\documentclass[journal]{IEEEtran}
\makeatletter
\def\ps@headings{%
\def\@oddhead{\mbox{}\scriptsize\rightmark \hfil \thepage}%
\def\@evenhead{\scriptsize\thepage \hfil \leftmark\mbox{}}%
\def\@oddfoot{}%
\def\@evenfoot{}}
\makeatother

\usepackage{fancyhdr}
\usepackage{graphicx}
\usepackage{subfigure}
\usepackage{amsmath}
\usepackage{epstopdf}
\usepackage{booktabs}
\usepackage[english]{babel}
\usepackage{amssymb}

\usepackage[ruled,vlined]{algorithm2e}
\usepackage{tipa}
\usepackage{lpic}
\usepackage{stfloats}
\usepackage{algorithmic}
\usepackage{url}
\usepackage{verbatim}
\usepackage{empheq}
\usepackage{amsthm}
\usepackage{cases}
\usepackage{multirow}
\usepackage{longtable}
\usepackage{tabu}

\ifCLASSINFOpdf

\else

\fi

\begin{document}

\title{SDN-Based Resource Management for Autonomous Vehicular Networks: A Multi-Access Edge Computing Approach}

\author{\IEEEauthorblockN{Haixia Peng,~\IEEEmembership{Student Member,~IEEE}, Qiang Ye,~\IEEEmembership{Member,~IEEE}, Xuemin (Sherman) Shen,~\IEEEmembership{Fellow,~IEEE}}

\thanks{
H. Peng, Q. Ye, and X. Shen are with the Department of Electrical and Computer Engineering, University of Waterloo, Waterloo, ON, Canada, N2L 3G1 (e-mail: \{h27peng, q6ye, sshen\} @uwaterloo.ca).
}}

\maketitle
\cfoot{\thepage}
\renewcommand{\headrulewidth}{0pt}
\renewcommand{\footrulewidth}{0pt}
\pagestyle{fancy}
\cfoot{\thepage}

\begin{abstract}
Enabling high-definition (HD)-map-assisted cooperative driving among autonomous vehicles (AVs) to improve the navigation safety faces technical challenges due to increased communication traffic volume for data dissemination and increased number of computing/storing tasks on AVs. In this article, a new architecture that combines multi-access edge computing (MEC) and software-defined networking (SDN) is proposed for flexible resource management and enhanced resource utilization. With MEC, the interworking of multiple wireless access technologies can be realized to exploit the diversity gain over a wide range of radio spectrum, and at the same time, computing/storing tasks of an AV are collaboratively processed by servers and other AVs. Moreover, by enabling SDN and network function virtualization (NFV) control modules at each cloud-computing and MEC server, an efficient resource allocation framework is proposed to enhance global resource sharing among different network infrastructures. A case study is presented to demonstrate the effectiveness of the proposed resource allocation framework.

\end{abstract}

\section*{Introduction}
\label{sec:Intro}

With the advanced artificial intelligence and automobile technologies, autonomous vehicles (AVs) have attracted great attention from both industry and academia. In recent years, a plethora of research activities and driving tests have been undergoing for AVs. It is expected that AVs will appear on the roads and even replace some manually driving vehicles in the near future \cite{human2018li}. Through combining a variety of techniques, such as radar, laser light, and computer vision, an AV with full driving automation senses environment and navigates without human actions or interventions \cite{autonomous2018su}. Existing works indicate that AVs have the potential to help solving traffic related problems, including avoiding the traffic accidents caused by human mistakes and reducing traffic congestion, energy consumption, and exhaust pollution \cite{kockelman2017assessment}.

Despite the potential advantages, the market perspective of AVs continues to face significant challenges. The autonomous navigation accuracy of an AV depends on the timeliness and granularity of its perceived and predicted road environment, which is constrained by the AV's sensing capability. Due to the deficiency in sensing accuracy, especially in road surroundings with bad weather, confusing traffic signals, and faded lanes, it is difficult to guarantee complete AV safety without human interventions. Existing studies indicate that enabling cooperative driving among AVs and providing high-definition (HD) maps to AVs emerges as a complementary technology to compensate the sensing deficiency and improve the AV safety \cite{sabuau2017optimal}. However, guaranteeing the performance of information interaction among AVs is challenging from the following aspects:
\begin{enumerate}
\item \textbf{Increased communication data traffic volume and computing/storing tasks:} Information from both onboard sensors and vehicle-to-vehicle/vehicle-to-infrastructure (V2V/V2I) interactions is required for cooperative driving, which increases the data traffic, such as sharing velocity and acceleration information between AVs. Moreover, building/updating the HD maps of a road environment, which are then shared with other AVs, leads to a large amount of data traffic in vehicular communication networks. To accommodate the surge of data traffic volume, an integration of different types of wireless communication access technologies is required. For example, the interworking between dedicated short range communications (DSRC) and cellular networks. Moreover, to support cooperative driving among AVs, the number of computing/storing tasks performed on some AVs increases to collect and process the information from both onboard sensors and from other vehicles. However, the computing and storing capabilities on each individual AV are limited to afford the increasing demands \cite{huang2018v2v};
\item \textbf{Heterogeneous quality of service (QoS) requirements:} Stringent delay requirement with guaranteed reliability is important to cooperative driving and safety-related applications. For example, a short response-delay from the vehicle traffic management system is required to maintain efficient cooperative driving. However, high throughput is required for transmitting high data traffic caused by some non-safety applications, such as HD map information and infotainment services. To guarantee the heterogeneous application QoS requirements, efficient QoS-oriented resource management is essential.
\end{enumerate}
How to support the increased traffic data in vehicular networks and the increased number of computing/storing tasks on AVs to reduce the response-delay for cooperative driving and deliver HD maps to AVs in time needs further research.

In this article, we introduce a new network architecture, which combines the approaches of multi-access edge computing (MEC) and software-defined networking (SDN) \cite{Zhang2017mobile}, to address the preceding challenges. Specifically, the proposed architecture has the following advantages:
\begin{enumerate}
\item Computing and storing resources are installed at the MEC servers to enable computing and storing capabilities at the edge of the core network, such that a short response-delay can be provided to cooperative driving, and HD maps can be cached in and processed by the MEC servers;
\item Through network function virtualization (NFV), efficient and dynamic computing and storing resource management is possible among different MEC servers. Computing/storing task balancing among MEC servers can be achieved by the NFV control module in the cloud-computing server, and the scalability of the architecture can be achieved;
\item SDN control modules in the cloud-computing/MEC servers separate the control plane from the data plane. Via MEC, multiple wireless access networks can interwork to support the increased traffic data, and various radio spectrum resources can be abstracted and reallocated to base stations (BSs) by the SDN control modules in MEC servers;
\item By combining the approaches of MEC and SDN, computing/storing resource allocation and bandwidth resource allocation can be jointly considered, and therefore to collaboratively satisfy AV delay requirements.
\end{enumerate}

The remainder of this article is organized as follows. First, the challenges in autonomous vehicular networks (AVNETs) are introduced and an AVNET architecture which incorporates both MEC and SDN technologies is proposed. Then, resource management schemes are presented, and a case study is provided to demonstrate the effectiveness of our proposed schemes. Finally, we discuss future research issues and draw concluding remarks.

\section*{An Autonomous Vehicular Network Architecture with MEC and SDN}
\label{sec:Autono}

In the following, we discuss research challenges in AVNETs, investigate how to apply MEC and SDN to deal with the challenges, and propose an AVNET architecture.

\subsection*{Challenges in AVNETs}
\label{subsec:Chall}

Existing studies indicate that enabling cooperative driving among AVs and providing real-time HD maps to AVs are important to improve AV safe navigation \cite{de2014network}. AVs with the same driving direction can be grouped into platoons/convoys for cooperative driving. In each platoon (or each convoy), neighboring AVs on one lane (or multiple lanes) move with a steady speed and keep a small steady inter-vehicle space \cite{sabuau2017optimal}. A leader vehicle is chosen to lead all the other AVs (referred to as member vehicles) within a platoon (or a convoy) to maintain the string stability \cite{fernandes2012platooning, peng2017vehicular}.

The performance of platoon (or convoy)-based driving varies in different road environments. Platooning can improve AV safe navigation in some highways or urban major avenues where vehicle density is relatively stable, whereas it may incur extra cost (e.g., time waste and fuel consumption) in urban areas due to constantly changing memberships in each platoon. On the other hand, HD maps can assist AVs for accurate accelerating or decelerating prior to obtaining the sensing information. Therefore, a promising AVNET scenario is a co-existence of HD-map-assisted cooperative driving and free driving, which can better adapt to different road environments. However, this new AVNET scenario poses technical challenges on both communication and computing.

From the communication perspective, delay-sensitive information needs to be shared among cooperative AVs. For example, to enable the cooperative adaptive cruise control for string stability, the speed and acceleration information of the leader vehicle and its preceding vehicle is required for each member vehicle in a platoon \cite{fernandes2012platooning}. In some communication-assisted platoon control schemes \cite{peng2017resource}, member vehicles are required to share their braking or leaving information with the leader vehicle, and then the leader vehicle makes accelerating or braking decisions for and shares the information with member vehicles. Moreover, static HD maps illustrating the static road environment (such as the lane lines and the surrounded buildings and trees) require a large volume of data traffic within the core network and data transmission to AVs. Likely, a single wireless access technology cannot fulfill the communication requirements, and how to improve the available resource utilization is critical, due to network resource scarcity.

From the computing perspective, high computing and storing capabilities are required by leader vehicles to process the aggregated information to make accelerating or braking decisions, and increasing computing/storing resources at each AV can be cost-ineffective. Also, efficiently handling dynamic information of AV speeds and traffic flow is necessary for timely updating dynamic HD maps\footnote{The dynamic HD map provides dynamic environment information to each AV for high localization precision, e.g., driving status information about the adjacent AVs.} at AVs. Due to the limited computing/storing resources of each AV, offloading the AV tasks with high computing/storing resource requirements to the remote cloud-computing server or to other AVs are two appealing ways. However, offloading all such tasks to the cloud-computing server can cause a significant burden on the core network and high response delay to AVs \cite{huang2018v2v}, whereas performing collaborative computing only among AVs with guaranteed QoS requirements is also difficult due to high vehicle mobility. To address the challenges from both computing and communication aspects in AVNETs, an effective solution relies on increasing the computing capability and integrating different access technologies\footnote{AVs are assumed to have multiple communication technologies interfaces to allow them access network through different BSs \cite{zheng2015dynamic}, such as roadside units (RSUs), Wi-Fi access points (APs), LTE base station e-Node B (eNB), and White-Fi infostations.} in AVNETs.

\subsection*{Proposed AVNET architecture}
\label{subsec:Why}

In this subsection, an SDN-enabled MEC architecture is proposed for the AVNET.

\begin{figure*}[htbp]
\centering
\includegraphics[height=4.2 in]{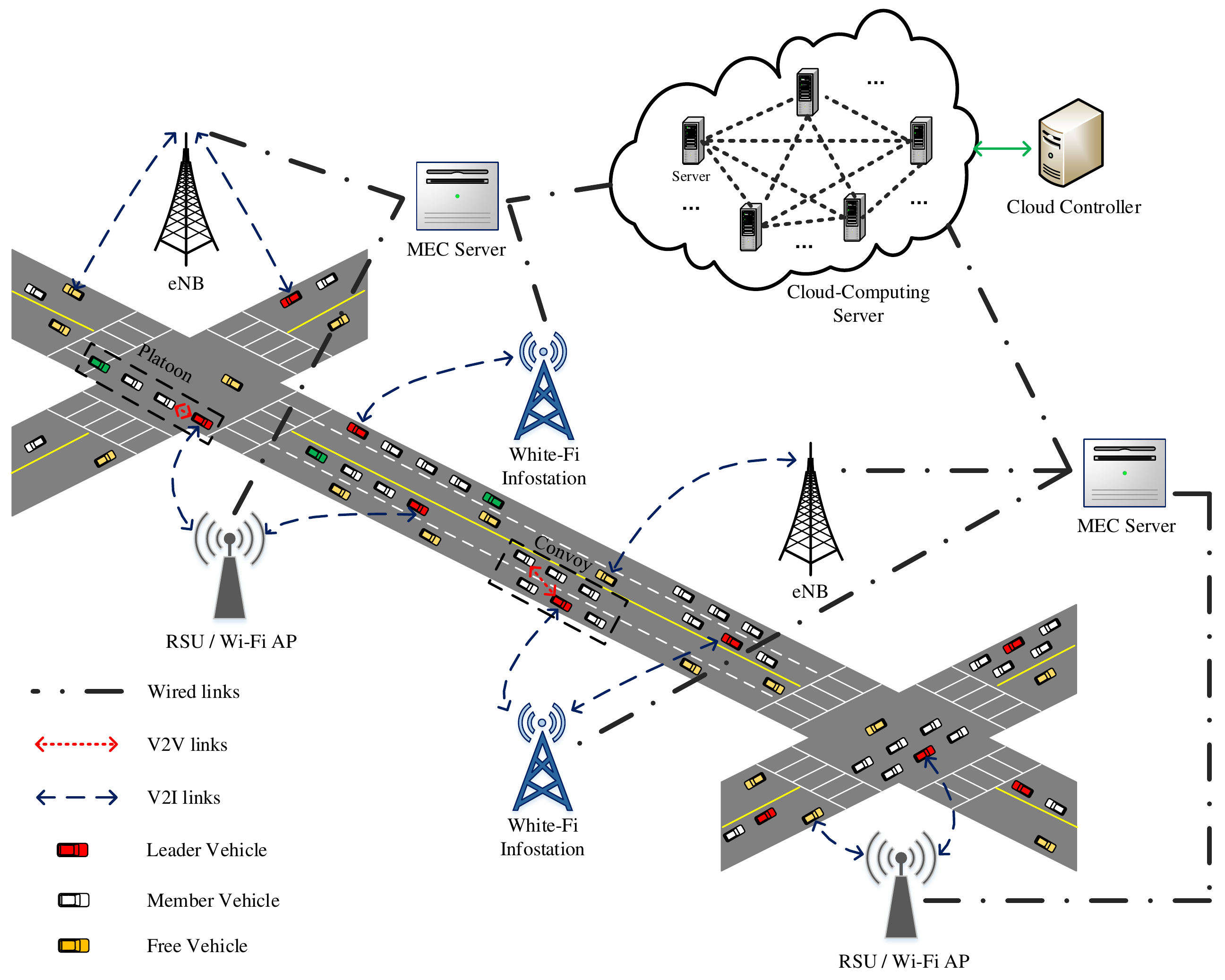}
\caption{An SDN-enabled MEC architecture.}
\label{fig:systemTo}
\end{figure*}

\textbf{\emph{An SDN-enabled MEC architecture:}} The proposed SDN-enabled MEC architecture is illustrated in Figure \ref{fig:systemTo}. Via moving computing and storing capabilities to MEC servers to form a two-tier server structure \cite{NFE-EVE, cheng2018agmenarx}, i.e., a cloud-computing server in the first tier and some MEC servers in the second tier, AV applications can be supported either by the cloud-computing server or MEC servers. Therefore, tasks with high computing/storing requirement of AVs, such as the computing tasks of leader vehicles for platooning management, can be offloaded to the MEC servers for quick responses to AVs by avoiding the data transfer between MEC servers and the cloud-computing server in the core network. Based on distinctive QoS requirements, delay-sensitive applications (e.g., the dynamic HD map management) are prioritized for registration in the MEC servers rather than delay-tolerant applications. As shown in Figure \ref{fig:systemTo}, a cloud-computing server, which consists of a group of servers, is placed at the core network. Moreover, a controller composed of NFV and SDN control modules is installed at the cloud-computing server (called as cloud controller for short) to centrally manage routing in the core network, computing/storing resources in the cloud-computing server, and idle computing/storing resources in MEC servers. The idle computing/storing resources of an MEC server is the residual resources that are available for handling tasks offloaded from other MEC servers.

To improve both resource utilization and network scalability, MEC servers are placed at the edge of the core network rather than at each BS, such that an MEC server can control the computing tasks for a large number of vehicles under the coverages of several BSs, and the enhanced service area of MEC server will better support high mobility of vehicles. The service area of an MEC server is defined as the total coverage area of BSs connected to it. Each MEC server forwards its state information, including the amount of idle computing/storing resources and QoS demands from different AV applications, to the core network after pre-analysis and pre-process (e.g, quantization). In such a way, the cloud controller can obtain computing/storing resource usage information from all MEC servers to centrally manage task migration among MEC servers. In order to allocate the local computing/storing resources, including computing/storing resources in AVs and in the MEC server, to different AV tasks meanwhile guaranteeing the heterogenous QoS requirements, a controller integrated of NFV and SDN control modules is installed at each MEC server \cite{huang2018v2v}, which is called as MEC controller for short. In addition, the MEC controller is in charge of integrating various bandwidth resources of different access networks to support the increasing data traffic transmission.

\begin{figure}[htbp]
\centering
\includegraphics[height=2.0 in]{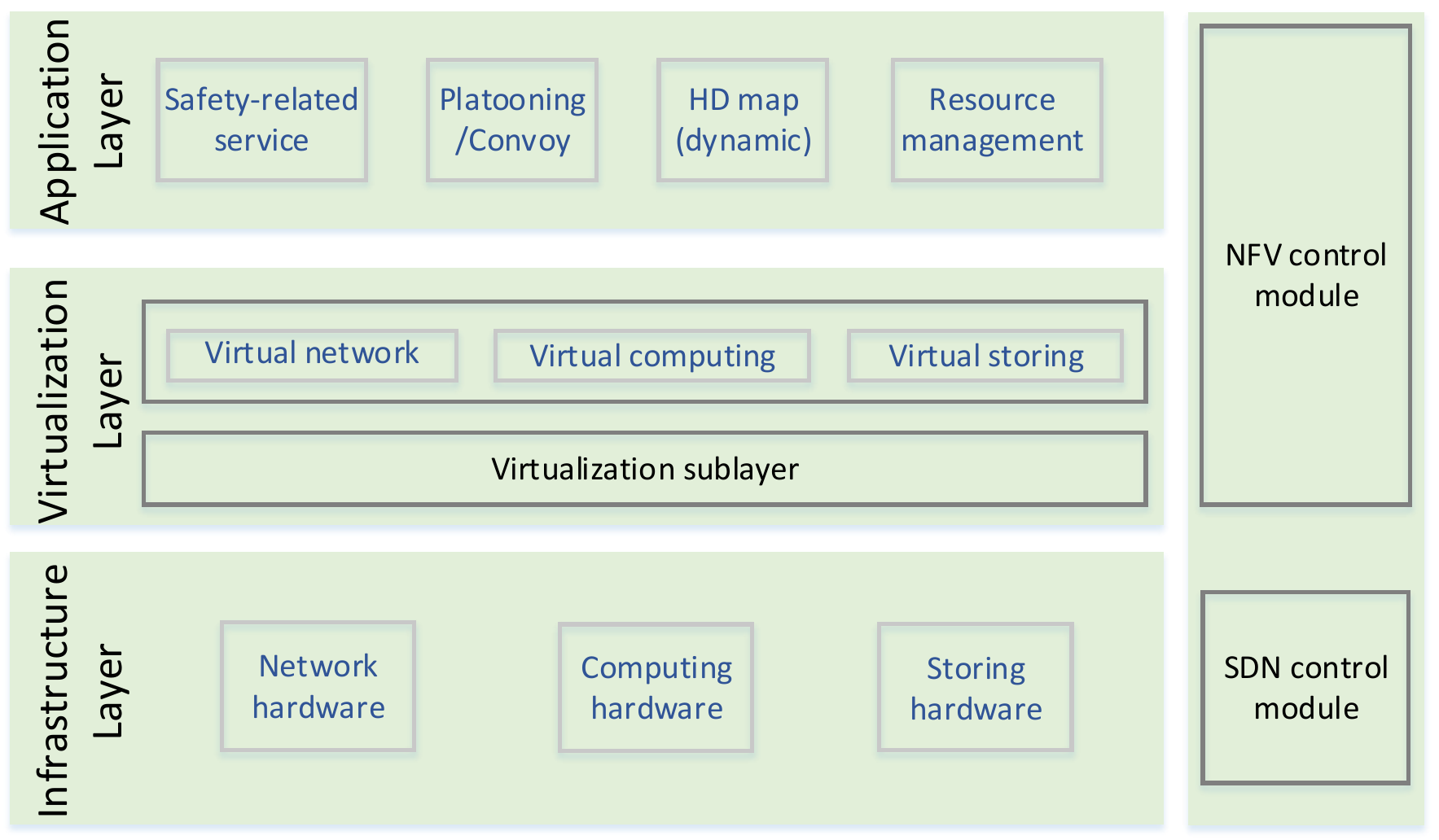}
\caption{Logically-layered structure for an MEC server.}
\label{fig:MEC}
\end{figure}

\textbf{\emph{A Logically-Layered Structure:}} To better illustrate the internal information exchange among different network components (functionalities), the proposed network architecture can be explained by using a logically-layered presentation from both the MEC and the cloud-computing perspectives. Since a logically-layered structure for an MEC server shares some common components with a cloud-computing server, we describe both structures separately and emphasize their differences. Each logically-layered structure is composed of an infrastructure layer, a virtualization layer, an application layer, and a separate control functionality, as shown in Figure \ref{fig:MEC} for an MEC server.

The infrastructure layer represents the underlying hardware, including computing hardware and memory hardware placed at AVs and at cloud-computing/MEC servers, and network hardware such as physical remote radio heads (RRHs) in BSs and baseband units (BBUs) deployed at MEC servers. Even though the computing hardware, memory hardware, and network hardware can be managed by the cloud controller or MEC controllers, how to improve their utilization to support the ever-increasing AV applications is challenging due to the regional distribution of hardware resources and the changing density of AVs. To address that, the resource virtualization technique is utilized in the proposed architecture to achieve resource programmability and enhance global resource utilization.

Underlying hardware resources are virtualized at the virtualization layer to make computing or storing environments independent of physical components. By doing that, virtual resources can be sliced and reallocated to different applications based on their QoS requirements by the cloud/MEC controllers, and each application or service function is decoupled from the physical components and run as software instances on the virtual resources. Therefore, one or more application services can be supported by one MEC server, and application can be flexibly added in or removed from the cloud-computing/MEC server without impacts on other applications. The decoupled applications or service functions are represented at the application layer. Considering heterogenous QoS requirements and available resources in MEC servers, delay-sensitive AV applications are prioritized for registration in the MEC servers, such as safety-related services, platooning/convoying, and dynamic HD map management. Delay-tolerant computing tasks (e.g., hot information caching, entertainment services, and static HD map management) are registered in the cloud-computing server.

The control functionality implemented in cloud/MEC controllers is composed of two modules, i.e., NFV control module and SDN control module. For both of the cloud controller and MEC controller, the NFV control module is responsible for resource virtualization, service function virtualization, and function placement, such as abstracting applications registered in the cloud-computing/MEC servers as different virtual service functions, and allocating virtual resources to each virtual service function. For example, to better utilize the computing and storing resources of the AVNET, the NFV control module in the cloud controller globally orchestrates the virtual computing/storing resources at the cloud-computing servers and the idle virtual computing/storing resources at each MEC server. In an MEC controller, the NFV control module locally orchestrates the virtual computing/storing resources at the MEC server and at each AV, and abstracts bandwidth resources and reallocates them to each connected BS, known as bandwidth slicing \cite{ye2018end}. Unless otherwise specified, bandwidth resources mentioned in this work are wireless bandwidth resources. Different from NFV control module, the SDN control module in each controller is responsible for centralized routing and forwarding management through abstracting the control functions from the data plane. The interaction between the control plane and the data plane is realized through the data-controller interface enabled by the OpenFlow protocol. Data flows going through the core network are under the control of the SDN control module in the cloud controller, and data flows among AVs and BSs are controlled by the SDN control module in MEC controller.

\section*{SDN-enabled Resource Management}
\label{sec:Autono}

Based on the proposed AVNET architecture with MEC and SDN in Figure \ref{fig:systemTo}, the increasing data traffic transmission and heterogenous QoS requirements can be supported by the integrated multiple access networks and the computing/storing resources installed at MEC servers. Due to the high cost of computing/storing resource installing and bandwidth resource scarcity, it is important to design efficient resource management schemes to improve the cost efficiency and resource utilization while guaranteeing the heterogeneous application QoS requirements for AVs. In this section, we investigate how to dynamically manage computing/storing resources among MEC servers and slice bandwidth resources among BSs.

\subsection*{Resource management schemes}
\label{subsec:ResMan}

For MEC servers with pre-allocated computing and storing resources, inter-MEC resource sharing is of paramount importance. Through migrating computing/storing tasks from one MEC server to other MEC servers, computing/storing resource utilization can be improved, where the tasks' processing results will be returned to the original MEC server to respond to the requester. Computing/storing tasks can also be migrated from one MEC server to a new MEC server based on the AV's moving direction, and the tasks' processing results will be directly delivered to each requester once the requester moves into the service area of the new MEC server. In the following, resource management schemes including computing and storing resource management and bandwidth management schemes are investigated.

Consider one cloud-computing server, $M$ MEC servers denoted by $M_i$, $i = 1,\ldots, M$, and $N$ AVs distributed over the entire AVNET. Each BS is connected to one of the $M$ MEC servers, and Wi-Fi/DSRC, White-Fi, and cellular technologies are applied to support AV applications. For MEC server $M_i$, the computing, storing, and bandwidth resources are denoted as $C_i^{\rm max}$, $S_i^{\rm max}$, and $B_i^{\rm max}$, respectively, where $B_i^{\rm max}$ is the total available bandwidth resources of the multiple radio access technologies for AVNETs. At time slot $t$, the resource allocated to AV $k$ is denoted by $D^{k}(t)=\{C^{k}(t),S^{k}(t),B^{k}(t)\}$, including the required amounts of computing resources $C^{k}(t)$ and storing resources $S^{k}(t)$ for processing its application requests, and bandwidth resources $B^{k}(t)$ for downlink transmissions. For AV $k$, we have $C^{k}(t) \geq 0$, $S^{k}(t) \geq 0$, and $B^{k}(t) > 0$, which means that not all AVs require computing and storing resources. Moreover, each of the required resources, i.e., $C^{k}(t)$, $S^{k}(t)$, and $B^{k}(t)$, can only be provided by one MEC server, and the three types of resources can be provided by the same MEC server or different MEC servers.

\emph{Computing and storing resource management:} Let $C_i^{k}(t)$ and $S_i^{k}(t)$ denote the amounts of computing and storing resources that MEC $M_i$ allocates to AV $k$ at time slot $t$. The computing resource utilization of MEC server $M_i$ is defined as the ratio of occupied resources over its total available computing resources, i.e., $\sum_{k=1}^{N_i^c(t)} \frac{C_i^{k}(t)}{C_i^{\rm max}}$. Similarly, the storing resource utilization is defined as $\sum_{k=1}^{N_i^s(t)} \frac{S_i^{k}(t)}{S_i^{\rm max}}$, where $N_i^c(t)$ and $N_i^s(t)$ are the numbers of AVs with $C_i^{k}(t)>0$ and $S_i^{k}(t)>0$, respectively. Due to the physically fixed computing/storing resources at each MEC server and the varying amount of computing/storing tasks generated by the regionally distributed moving AVs, each MEC server processing the computing/storing tasks of AVs within its service area can cause the MEC server to be task overloaded or underloaded. To mitigate the imbalanced task scheduling, computing/storing tasks can be migrated among MEC servers and, therefore increasing the computing/storing resource utilization. However, it causes task migration cost, such as consuming wired bandwidth resources, and adds extra delay to AVs.

\begin{figure}[htbp]
\centering
\includegraphics[height=2.20 in]{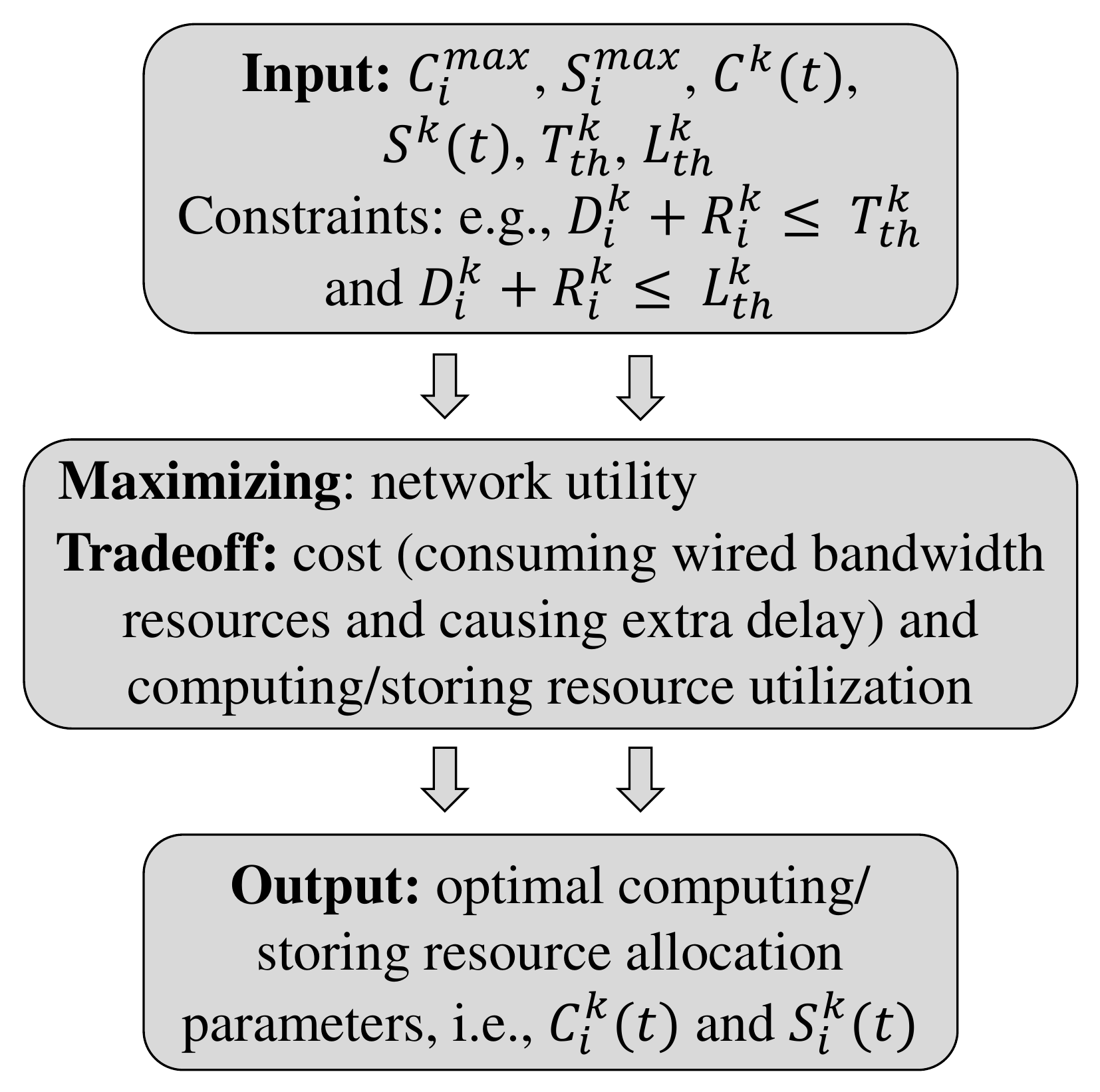}
\caption{Diagramming for formulating optimization problem of computing/storing resources.}
\label{fig:Opti_pro}
\end{figure}

In order to obtain optimal computing and storing resource allocation while balancing the tradeoff between increasing the computing/storing resource utilization and reducing the task migration cost, a maximization problem can be formulated based on the optimization framework shown in Figure \ref{fig:Opti_pro}. The objective is to maximize the network utility which is defined as the summation of utilities of each individual MEC server. The utility of an MEC server allocating computing/storing resources to AVs is defined with the consideration of computing/storing resource utilization and task migration cost. The input of the formulated problem includes $C_i^{\rm max}$, $S_i^{\rm max}$, $C^{k}(t)$, $S^{k}(t)$, $T_{th}^k$, and $L_{th}^k$. $T_{th}^k$ is a downlink response delay threshold, which is used to guarantee that AV $k$ (either generating delay-sensitive requests or delay-tolerate requests) receives the response before it moves out consideration of the service area of the MEC server that provides computing/storing resources to it. $L_{th}^k$ is a latency threshold and is used to guarantee the delay requirement of AV $k$ generating a delay-sensitive request. Let $D_i^k$ be the task processing delay, i.e., the time interval from the time that AV $k$'s computing/storing task is received by MEC server $M_i$ until the MEC server finishes processing this task. When formulating the maximization problem, the following constraints should be considered: 1) $D_i^k+ R_i^k\leq T_{th}^k$ for AVs that either generate delay-sensitive requests or delay-tolerate requests, and $D_i^k + R_i^k \leq L_{th}^k$ for AVs that generate delay-sensitive requests, where $R_i^k$ is the downlink transmission delay from MEC server $M_i$ to AV $k$ through a BS, i.e., the time interval from the time that a packet reaches the transmission queue of MEC server $M_i$, until the time instant the packet is received by AV $k$ through a BS; 2) computing/storing resource constraints, such as $\Sigma_{i=1}^M C_i^{k}(t) = C^{k}(t)$ and $\Sigma_{i=1}^M S_i^{k}(t)= S^{k}(t)$. By solving the formulated maximization problem, the optimal computing and storing resource allocation can be obtained, $C_i^{k}(t)$ and $S_i^{k}(t)$.

\emph{Bandwidth management:} Due to the large service area of each MEC server, bandwidth reuse is considered among the BSs connected to the same MEC server. With the consideration of different BS coverages (RSUs/Wi-Fi APs, White-Fi infostations, and eNBs) from different wireless access technologies, AVs can choose to associate with BSs providing different levels of communication qualities (e.g., transmission rates). To improve bandwidth resource utilization, BSs can reuse bandwidth resources with acceptable inter-cell interference. Therefore, the goal of bandwidth slicing is to determine the set of optimal slicing ratios for different BSs, such that the aggregate network utility is maximized, and the heterogenous application QoS requirements are satisfied.

Taking BS $j$ and AV $k$ under the service area of MEC server $M_i$ as an example, let $\gamma_{j}^{k}(t)$ denote the achievable downlink transmission rate when AV $k$ is associated with BS $j$ at time slot $t$. The utility of AV $k$ associated with BS $j$ is defined as a concave function of $\gamma_{j}^{k}(t)$, e.g., a logarithmic function, and the aggregate network utility is defined as the summation of utilities of each individual AV. Then, a network utility maximization problem is formulated, in which a two-level resource allocation is considered: 1) slicing the total bandwidth resources $B_i^{\rm max}$ into small resource slices, $\{\beta_j|j=1,2,\cdots,I_i(t)\}$, where $\sum_{j=1}^{I_i(t)} \beta_j = 1$ and $I_i(t)$ is the number of BSs within the service area of MEC server $M_i$ at time slot $t$; 2) partitioning the sliced resources to different AVs under the coverage of each BS. Constraints should be considered in this formulated problem include: 1) $D_i^k + R_i^k\leq T_{th}^k$ and $D_i^k + R_i^k \leq L_{th}^k$; 2) $\gamma_{j}^{k}(t) \geq \widehat{\gamma}^k$, where $\widehat{\gamma}^k$ is defined as a transmission rate threshold for AVs that generate delay-tolerant requests; 3) bandwidth resource constraints, $B_i^{\rm max}$.

The latency constraints for AV $k$ reflect the coupling relation between the two problems formulated for computing/storing resources and bandwidth resources. Thus, these two problems have to be jointly solved, and the obtained optimal computing/storing resource allocation results and bandwidth resource allocation results maximize the network utility while collaboratively satisfying the delay requirement for each AV.

\subsection*{Case study}
\label{subsec:case}

\begin{figure}[htbp]
\centering
\includegraphics[height=1.70 in]{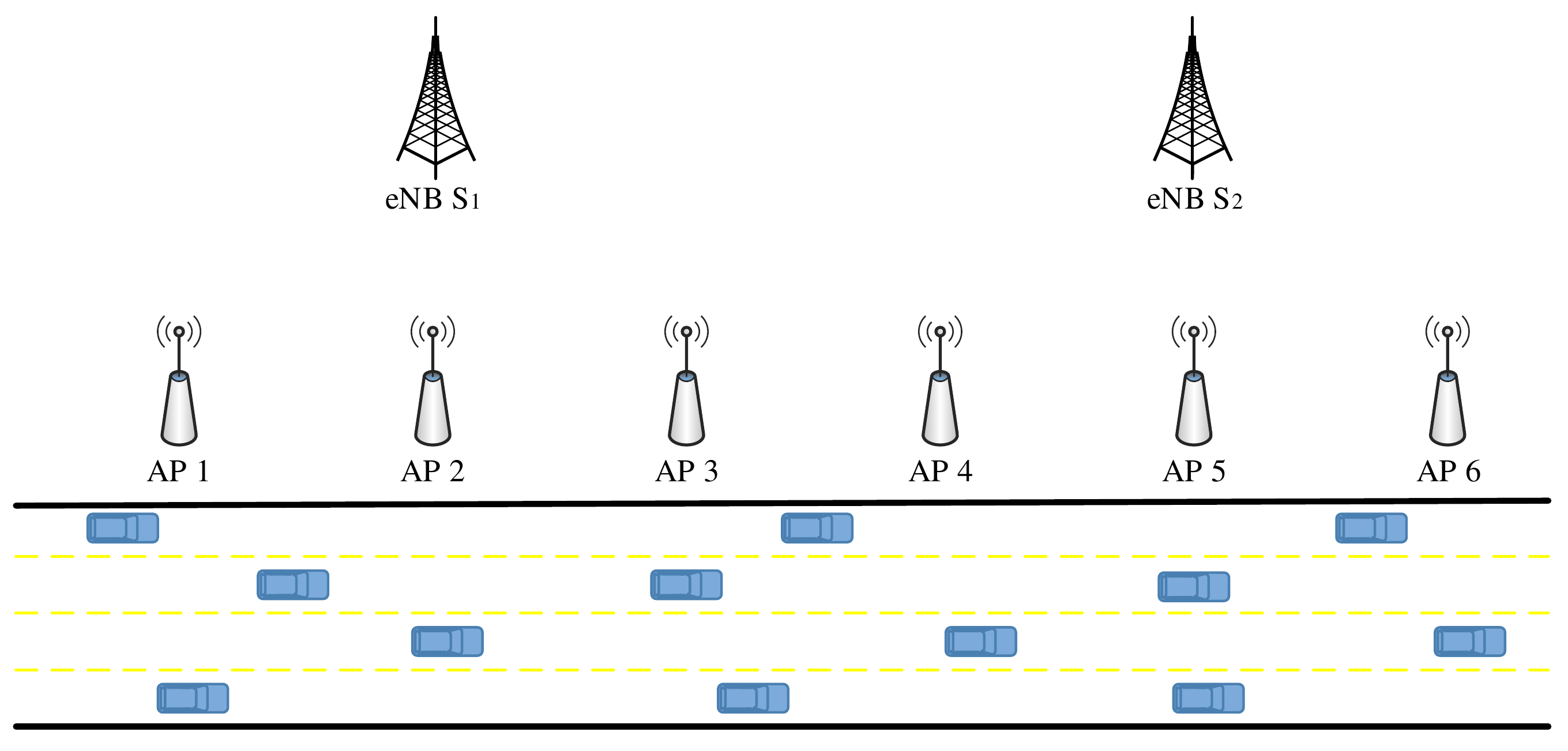}
\caption{A network scenario under consideration.}
\label{fig:simple}
\end{figure}

In this subsection, a case study is presented to demonstrate the effectiveness of our resource management scheme over a simple network scenario, where cellular and Wi-Fi communication technologies are employed to support AV applications. As shown in Figure \ref{fig:simple}, $2$ eNBs (eNB $S_1$ and eNB $S_2$) and $6$ Wi-Fi APs (AP $1$ to AP $6$) are distributed over one side of a one-way straight road with four lanes. Locations of the eNBs and Wi-Fi APs are fixed. Transmit power of each eNB is set to be $40$ dBm (i.e., $10$ watts) to ensure a maximum communication range of $600$ m, which fully covers the AVs on the road, whereas the transmit power of a Wi-Fi AP is $28.45$ dBm (i.e., $0.7$ watts) with the communication range of $180$ m and the coverage rate of Wi-Fi APs is less than or equal to $1$. The coverage rate of a type of BS is defined as the probability that an AV is within the coverage of the BS. The AV density over four lanes, i.e., the number of AVs in the four lanes per meter, is assumed to vary within range of $[0.12, 0.40]$AV/m. In our simulation, the number of AVs per 100 meters in the four lanes is randomly chosen from $12-40$. For each AV, only one type of delay-sensitive applications (transmitting safety-related information) or delay-tolerant applications (downloading HD map information) is considered in each time slot, where the probability that the AV generates a delay-sensitive application request is set to $0.8$. To balance the bandwidth resource utilization and inter-cell interference, bandwidth reuse is only considered between two Wi-Fi APs and between one eNB and one Wi-Fi AP (only the Wi-Fi APs having no overlapping coverage area with the eNB can reuse the bandwidth allocated to the eNBs). Thus, the total available bandwidth is sliced into three slices allocated to eNB $S_1$, eNB $S_2$, and Wi-Fi APs with the slicing ratios $\beta_1$, $\beta_2$, and $\beta_w$, respectively. For example, AVs within the coverages of APs $4$, $5$, and $6$ can reuse the bandwidth resources $\beta_w B_{i}^{\rm max}$. Since the coverage areas of APs $4$, $5$, and $6$ have no overlapping with that of eNB $S_1$, they can also reuse the amount of bandwidth resources $\beta_1 B_{i}^{\rm max}$. Other important simulation parameters are listed in Table \ref{table:parameters}.

\begin{table}[htbp]
\setlength{\belowcaptionskip}{5pt}
 \caption{\label{table:parameters} Parameters values}
 \centering
 \begin{tabular}{p{5.4cm}||p{2.2cm}}
  \toprule
  \textbf{Parameter} & \textbf{Value}\\
  \midrule
The minimum inter-vehicle distance & $5$ m \\
Propagation model for the eNB & $-30-35 {\rm log}_{_{10}}(d)$ \\
Propagation model for the Wi-Fi AP & $-40-35 {\rm log}_{_{10}}(d)$ \\
MAC efficiency factor of Wi-Fi technology & $0.8$ \\
Background noise power & -104 dBm \\
Aggregate bandwidth resources ($B_i^{\rm max}$) from BSs and Wi-Fi APs under MEC $i$ & $25$ MHz \\
HD map packet arrival rate & $20$ packet/s \\
HD map packet size & $9000$ bits \\
Safety-related packet arrival rate & $4$ packet/s \\
Safety-related packet size & 1048 bits \\
Safety-related packet delay bound & $100$ ms \\
Delay bound violation probability & $10^{-3}$ \\
  \bottomrule
 \end{tabular}
\end{table}

\begin{figure}[htbp]
\centering
\subfigure[]{
\label{fig:utility}
\includegraphics[width=0.45\textwidth]{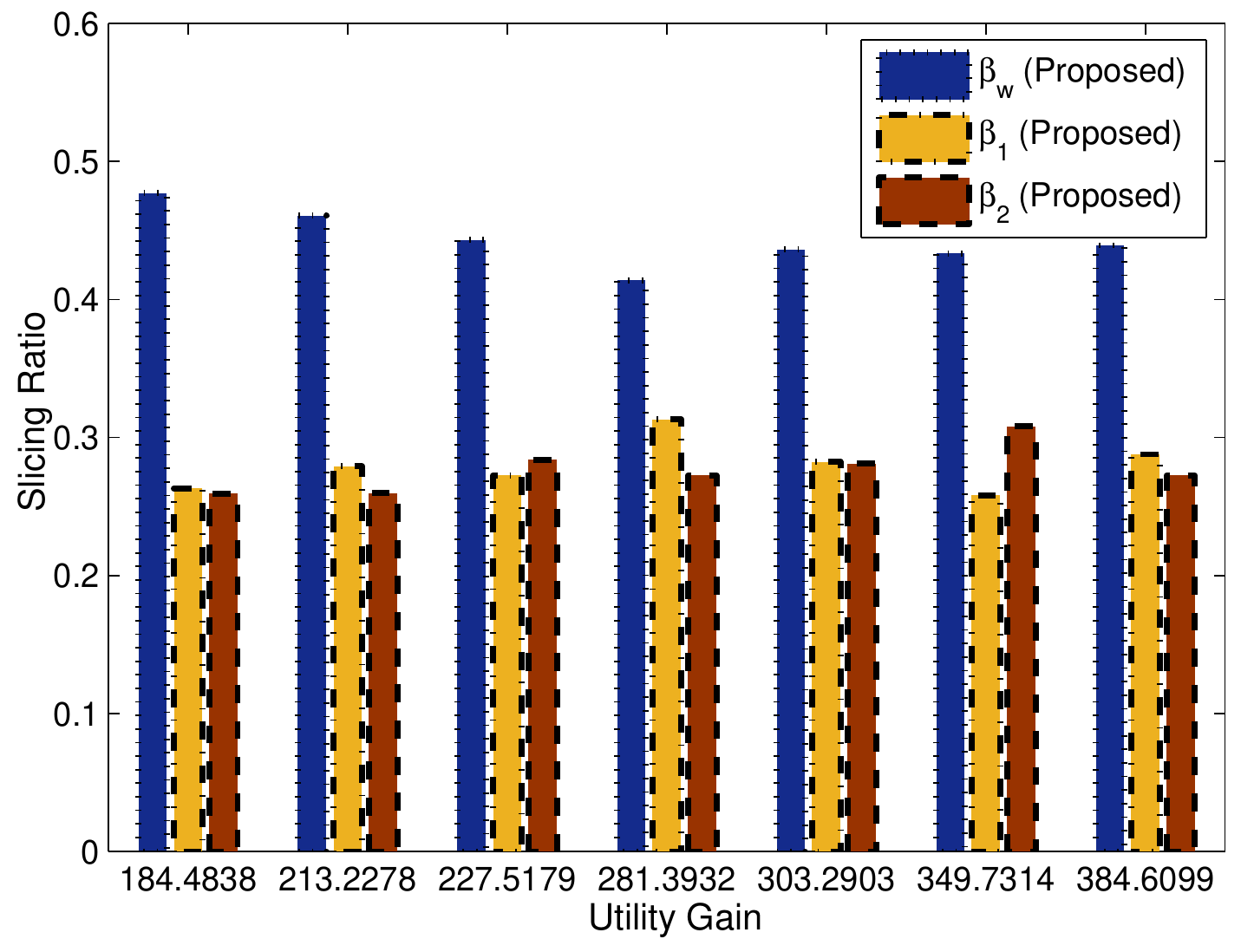}}
\subfigure[]{
\label{fig:association1}
\includegraphics[width=0.45\textwidth]{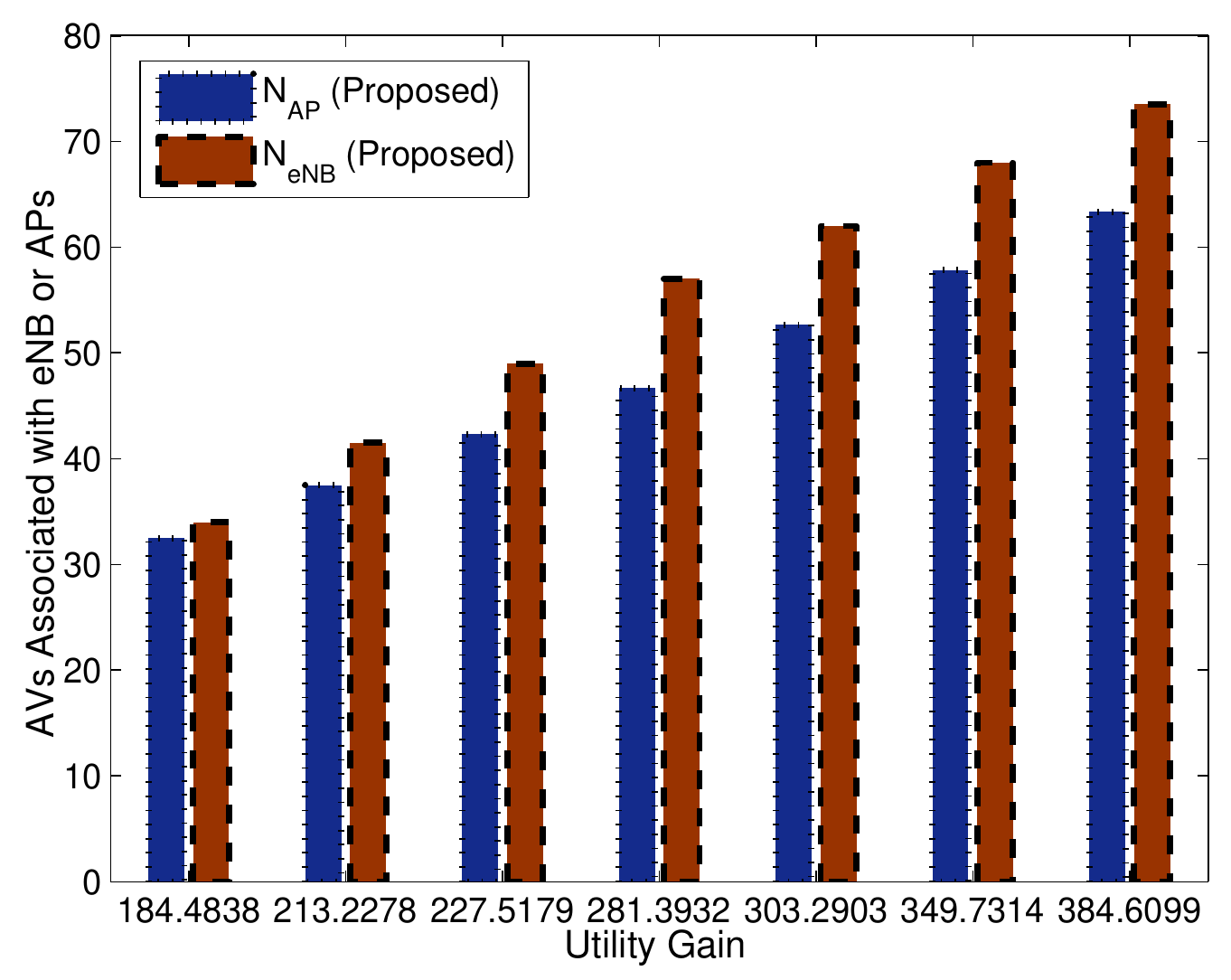}}
\subfigure[]{
\label{fig:association2}
\includegraphics[width=0.45\textwidth]{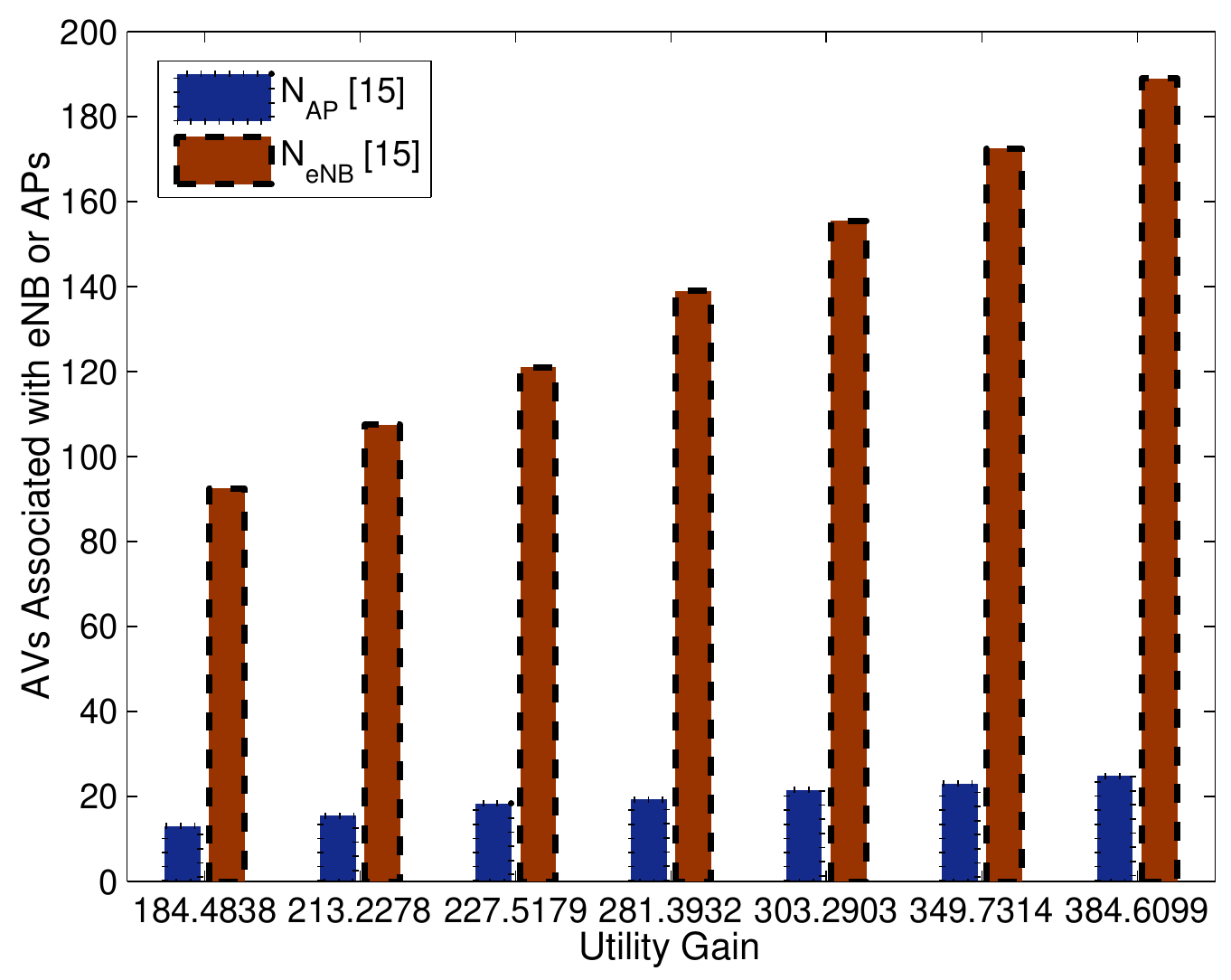}}
\caption{Bandwidth slicing ratios and average numbers of AVs associated with an eNB/AP under different utility gains.}
\label{fig:slicing}
\end{figure}

A logarithmic utility function is considered. For AV $k$ associated with BS $j$ (e.g., AP $4$), the utility function is defined as ${\rm log}(B_i^{\rm max} (\beta_1 + \beta_w)f_4^k \gamma_4^k)$, where $f_4^k$ is the fraction of bandwidth resources allocated to AV $k$ from Wi-Fi AP $4$ and $\gamma_4^k$ is the spectrum efficiency from Wi-Fi AP $4$. As discussed precedingly, an optimization problem that maximizes the network utility is formulated and certain approximation methods can be applied to solve this problem to obtain the optimal bandwidth slicing ratios $\{\beta_1^*, \beta_2^*, \beta_w^*\}$ and vehicle-BS association patterns. To demonstrate the efficiency of the proposed bandwidth slicing scheme, we compare our proposed scheme with the max-SINR scheme, proposed in \cite{liang2016virtual}, where no bandwidth slicing is enabled, and the end device associates with the BS providing the highest SINR level. Under scenarios with different vehicle densities, the bandwidth slicing ratios are dynamically adjusted in our proposed scheme, as shown in subfigure \ref{fig:utility}, where the horizontal axis represents the utility gain of our proposed scheme over the max-SINR scheme without bandwidth slicing under different vehicle densities, i.e., $\{0.12, 0.14, 0.16, 0.18, 0.20, 0.22, 0.24\}$AV/m. Through adjusting the bandwidth slicing ratios, the amount of bandwidth resources reused among Wi-Fi APs is optimized to improve the spectrum utilization while guaranteeing the heterogenous QoS requirements for different AV applications. Moreover, by simultaneously optimizing the bandwidth slicing ratios and the vehicle-BS association patterns, the proposed scheme balances the data traffic load among BSs compared with the max-SINR scheme, as shown in subfigures \ref{fig:association1} and \ref{fig:association2}, where $N_{\rm AP}$ and $N_{\rm eNB}$ are the average numbers of AVs associated with a Wi-Fi AP and an eNB, respectively. 
That is because for the max-SINR scheme, the bandwidth resources allocated to each BS are fixed, and most of the AVs within the coverage areas of APs choose to connect to eNB $S_1$ or eNB $S_2$ rather than APs due to higher received signal-to-interference and noise ratio (SINR) from eNBs. In our proposed bandwidth slicing framework, the numbers of AVs associated with different BSs are balanced, as shown in subfigure \ref{fig:association1}, through adjusting the bandwidth slicing ratios.

\section*{Open Research Issues}
\label{sec:Open}

Combining MEC with SDN to support the increasing data traffic volume while guaranteeing heterogeneous QoS requirements for different services in AVNETs, is still in its infancy. In this section, some open research issues are introduced.

\textbf{MEC deployment:} Basically, MEC servers should be placed at the edge of the core network close to users. In our proposed resource management framework, we consider that MEC servers are placed at the edge of the core network to maintain two-hop wireless transmissions between an AV and an MEC server. Placing MEC servers close to BSs reduces the computing task response delay, but increases the computing server deployment cost. Therefore, how to place MEC servers and how much computing and storing resources should be allocated to each MEC server need to be investigated for the MEC deployment problem. A simple method to deploy MEC servers is based on local service requirements to balance the placement cost with AVs' application QoS requirements. Furthermore, the MEC deployment and dynamic resource management should be jointly designed by considering service demand fluctuations due to the high AV mobility, vehicular traffic variations,, and increasingly diversified applications.

\textbf{Task offloading optimization:} Since computing/storing resources on each MEC server are limited and task migration from one MEC server to another incurs extra cost, the amount of tasks allowed to be registered and processed in MEC servers should be constrained. Therefore, designing a proper task offloading criterion is necessary to maximize the computing/storing resource utilization under the constraints of task migration costs. For the proposed framework, we determine where to register AV applications based on the application types, i.e., only safety-related applications are registered in the MEC servers. However, other performance criteria should also be taken into consideration to optimize the offloading decision. For example, delay requirements for each computing task. Therefore, given the amount of resources on each MEC server, how to design appropriate criteria for task offloading to balance the tradeoff between QoS satisfaction and minimizing the offloading cost is a challenging research issue.

\textbf{Bandwidth management for uplinks:} Different from downlink transmissions where BSs are at fixed locations, bandwidth allocation for uplink transmissions is more complex due to the following challenges: First, due to high AV mobility, the inter-cell interference changes dynamically and is difficult to be characterized; Second, it is inefficient for each vehicle to collect information from all neighboring vehicles to achieve local centralized control due to the highly dynamic network topology. To overcome these challenges, vehicle trajectory prediction schemes and distributed control methods can be applied when managing bandwidth for uplink transmissions.

\textbf{Fairness:} With the SDN control module, multiple access networks can be integrated to support AV applications. How to achieve fairness in selecting different wireless access technologies from end devices is an important research issue, where a proper fairness metric is required. Appropriate revenue models among network operators can be considered for designing a fair network selection and resource allocation scheme in terms of maximizing the revenue for each individual operator. From the end device perspective, an appropriate prioritization scheme is necessary among different AV applications, so that fairness among AVs can be well balanced while guaranteing the QoS satisfaction.

\textbf{Security and privacy:} How to ensure a secure communication among AVs is a key research issue. Since the accelerating or braking decisions from communication-assisted AVs are based on the collected information via V2V and V2I communications, security attacks on communication channels and sensor tampering may result in driving safety issues. Due to the MEC controllers, the privacy of applications registered in MEC servers can be improved through local communications. However, the MEC servers or cloud-computing servers can become the major targets of hackers, from which the attacks indirectly cause driving safety issues and result in serious privacy invasion. Moreover, exchanging individual vehicle information is required to support cooperative driving among AVs. How to ensure identity privacy, data privacy, and location privacy is essential to stimulate the cooperative driving among AVs. To deal with these security and privacy issues, potential solutions include identity authentication for communications, access control at MEC and cloud-computing servers, and trust management from AVs and servers.

\section*{Conclusion}
\label{sec:Conclu}

In this article, we have proposed a new networking architecture to enhance cooperative driving among AVs with increasingly intensified computing and communication requirements. By applying MEC in AVNETs, computing and storing resources are moved to the edge of the core network and AVs access the network via different wireless access technologies. To enhance the computing/storing and bandwidth resource utilization, a resource management framework is proposed under the new network architecture. Through centrally managing the computing/storing resources on cloud-computing and MEC servers, task load balancing among MEC servers can be enhanced. In addition, the SDN and NFV control modules at each MEC controller are also in charge of slicing the bandwidth resources of different access technologies among heterogeneous BSs to improve the bandwidth utilization. A case study has been conducted to demonstrate the effectiveness of the proposed resource management schemes. Some important open research issues related to the proposed architecture are also discussed.

\bibliographystyle{IEEEtran}
\bibliography{heterogeneousdriving}

\begin{thebibliography}{10}
\providecommand{\url}[1]{#1}
\csname url@samestyle\endcsname
\providecommand{\newblock}{\relax}
\providecommand{\bibinfo}[2]{#2}
\providecommand{\BIBentrySTDinterwordspacing}{\spaceskip=0pt\relax}
\providecommand{\BIBentryALTinterwordstretchfactor}{4}
\providecommand{\BIBentryALTinterwordspacing}{\spaceskip=\fontdimen2\font plus
\BIBentryALTinterwordstretchfactor\fontdimen3\font minus
  \fontdimen4\font\relax}
\providecommand{\BIBforeignlanguage}[2]{{%
\expandafter\ifx\csname l@#1\endcsname\relax
\typeout{** WARNING: IEEEtran.bst: No hyphenation pattern has been}%
\typeout{** loaded for the language `#1'. Using the pattern for}%
\typeout{** the default language instead.}%
\else
\language=\csname l@#1\endcsname
\fi
#2}}
\providecommand{\BIBdecl}{\relax}
\BIBdecl

\bibitem{human2018li}
L.~Li, K.~Ota, and M.~Dong, ``Human-like driving: Empirical decision-making
  system for autonomous vehicles,'' \emph{IEEE Trans. Veh. Technol.}, to
  appear.

\bibitem{autonomous2018su}
J.~Su, J.~Wu, P.~Cheng, and J.~Chen, ``Autonomous vehicle control through the
  dynamics and controller learning,'' \emph{IEEE Trans. Veh. Technol.}, to
  appear.

\bibitem{kockelman2017assessment}
\BIBentryALTinterwordspacing
K.~Kockelman, G.~Sharon \emph{et~al.}, ``An assessment of autonomous vehicles:
  traffic impacts and infrastructure needs - final report,'' University of
  Texas at Austin. Center for Transportation Research, Tech. Rep.
  FHWA/TX-17/0-6847-1, 2017. [Online]. Available:
  \url{https://library.ctr.utexas.edu/ctr-publications/0-6847-1.pdf}
\BIBentrySTDinterwordspacing

\bibitem{sabuau2017optimal}
{\c{S}}.~Sab{\u{a}}u, C.~Oar{\u{a}}, S.~Warnick, and A.~Jadbabaie, ``Optimal
  distributed control for platooning via sparse coprime factorizations,''
  \emph{IEEE Trans. Auto. Control}, vol.~62, no.~1, pp. 305--320, Jan. 2017.

\bibitem{huang2018v2v}
C.~Huang, M.~Chiang, D.~Dao, W.~Su, S.~Xu, and H.~Zhou, ``{V2V} data offloading
  for cellular network based on the software defined network {(SDN)} inside
  mobile edge computing {(MEC)} architecture,'' \emph{IEEE Access}, vol.~6, pp.
  17\,741--17\,755, Mar. 2018.

\bibitem{Zhang2017mobile}
K.~Zhang, Y.~Mao, S.~Leng, Y.~He, and Y.~Zhang, ``Mobile-edge computing for
  vehicular networks a promising network paradigm with predictive
  off-loading,'' \emph{IEEE Veh. Technol. Mag.}, vol.~12, no.~2, pp. 36--44,
  Jun. 2017.

\bibitem{de2014network}
\BIBentryALTinterwordspacing
A.~Fortelle, X.~Qian, S.~Diemer, J.~Gr{\'e}goire, F.~Moutarde, S.~Bonnabel,
  A.~Marjovi, A.~Martinoli, I.~Llatser, A.~Festag, and K.~Sjoberg, ``Network of
  automated vehicles: the {AutoNet} 2030 vision,'' in \emph{ITS World Congr.},
  2014. [Online]. Available:
  \url{https://hal-mines-paristech.archives-ouvertes.fr/hal-01063484/document}
\BIBentrySTDinterwordspacing

\bibitem{fernandes2012platooning}
P.~Fernandes and U.~Nunes, ``Platooning with {IVC}-enabled autonomous vehicles:
  Strategies to mitigate communication delays, improve safety and traffic
  flow,'' \emph{IEEE Trans. Intell. Transp. Syst.}, vol.~13, no.~1, pp.
  91--106, Mar. 2012.

\bibitem{peng2017vehicular}
H.~Peng, L.~Liang, X.~Shen, and G.~Y. Li, ``Vehicular communications: A network
  layer perspective,'' \emph{IEEE Trans. Veh. Technol.}, to appear.

\bibitem{peng2017resource}
H.~Peng, D.~Li, Q.~Ye, K.~Abboud, H.~Zhao, W.~Zhuang, and X.~Shen, ``Resource
  allocation for cellular-based inter-vehicle communications in autonomous
  multiplatoons,'' \emph{IEEE Trans. Veh. Technol.}, vol.~66, no.~12, pp.
  11\,249--11\,263, Dec. 2017.

\bibitem{zheng2015dynamic}
Q.~Zheng, K.~Zheng, L.~Sun, and V.~Leung, ``Dynamic performance analysis of
  uplink transmission in cluster-based heterogeneous vehicular networks,''
  \emph{IEEE Trans. Veh. Technol.}, vol.~64, no.~12, pp. 5584--5595, Dec. 2015.

\bibitem{NFE-EVE}
\BIBentryALTinterwordspacing
\emph{\BIBforeignlanguage{EN}{{ETSI GS NFV-EVE 005 V1.1.1 (2015-12)}}},
  European Telecommunications Standards Institute Std. gs nfv-eve 005, Dec.
  2015. [Online]. Available:
  \url{https://www.etsi.org/deliver/etsi_gs/NFV-EVE/001_099/005/01.01.01_60/gs_NFV-EVE005v010101p.pdf}
\BIBentrySTDinterwordspacing

\bibitem{cheng2018agmenarx}
N.~Cheng, W.~Xu, W.~Shi, Y.~Zhou, N.~Lu, H.~Zhou, and X.~Shen, ``Air-ground
  integrated mobile edge networks: Architecture, challenges and
  opportunities,'' \emph{IEEE Commun. Mag.}, to appear.

\bibitem{ye2018end}
Q.~Ye, J.~Li, K.~Qu, W.~Zhuang, X.~Shen, and X.~Li, ``{End-to-End} quality of
  service in {5G} networks: Examining the effectiveness of a network slicing
  framework,'' \emph{IEEE Veh. Technol. Mag.}, vol.~13, no.~2, pp. 65--74, Jun.
  2018.

\bibitem{liang2016virtual}
C.~Liang, R.~Yu, H.~Yao, and Z.~Han, ``Virtual resource allocation in
  information-centric wireless networks with virtualization,'' \emph{IEEE
  Trans. Veh. Technol.}, vol.~65, no.~12, pp. 9902--9914, Dec. 2016.

\end{thebibliography}

\ifCLASSOPTIONcaptionsoff
  \newpage
\fi

\end{document}